\documentclass[12pt]{article}
\usepackage{graphicx}
\usepackage{amsmath,amssymb,amsfonts,multirow}
\usepackage{mathrsfs}
\usepackage{latexsym}
\usepackage{bm}
\usepackage{bbm}
\usepackage{cite}
\usepackage{subfigure}
\usepackage{mathrsfs}
\usepackage{indentfirst}
\usepackage{amsmath}
\usepackage{amssymb}
\usepackage{cases}
\usepackage{mathrsfs}
\usepackage{setspace}

\usepackage[colorlinks]{hyperref}
\hypersetup{citecolor=blue,linkcolor=blue,urlcolor=blue}

\begin{document}

\title{\Large\bf
A Pyramid Scheme Model Based on ``Consumer Rebate" Frauds}

\vspace{0.3truecm}
\author{Yong Shi\footnote{Email: yshi@ucas.ac.cn}, \  Bo Li\footnote{Email:libo312@mails.ucas.edu.cn}
  \ and Wen Long\footnote{Email: longwen@ucas.ac.cn, corresponding author}
 \\  \\
  \textit{School of Economics and Management,}\\
  \textit{University of Chinese Academy of Sciences, Beijing, 100190, China}\\
  \textit{Research Center on Fictitious Economy and Data Science},\\
  \textit{Key Laboratory of Big Data Mining and Knowledge Management,}\\
  \textit{Chinese Academy of Sciences, Beijing, 100190, China}
}

\date{\today}

\maketitle
\begin{abstract}
There are various types of pyramid schemes which have inflicted or are inflicting losses on many people in the world. We propose a pyramid scheme model which has the principal characters of many pyramid schemes appeared in recent years: promising high returns, rewarding the participants recruiting the next generation of participants, and the organizer will take all the money away when he finds the money from the new participants is not enough to pay the previous participants interest and rewards. We assume the pyramid scheme carries on in the tree network, ER random network, SW small-world network or BA scale-free network respectively, then give the analytical results of how many generations the pyramid scheme can last in these cases. We also use our model to analyse a pyramid scheme in the real world and we find the connections between participants in the pyramid scheme may constitute a SW small-world network.
\end{abstract}

\newpage

\section{Introduction}
The column Topics in Focus of China Central Television (CCTV), one of the most-watched shows on CCTV, exposed a so-called ``consumption rebate" platform
named ``RenRenGongYi" on June 18, 2017 which actually was a pyramid scheme\footnote{Topics in Focus on June 18, 2017 can be viewed at http://tv.cctv.com/2017/06/18/VIDE8gtfpFkpiB\\ v2QYnNuGIF170618.shtml.}. The alleged operation pattern of ``RenRenGongYi" was to let franchisees give the fixed proportion of customers' consumption to the platform to form a fund pool, such as 24\%, 12\% or 6\%, then the platform return the money to customers and
franchisees by instalments. But most transactions are fabricated in ``RenRenGongYi", and the platform in fact lured the participants by promising high returns to invest in the fund pool and rewarded the participants who attracted the next generation of participants. Franchisees may be mostly fictitious because most transactions are fabricated, and the participants of the pyramid scheme are mainly consumers. For example, If a participant fabricates a 100 yuan consumption
(the participant was both the franchisee and the consumer) and gave 24 (or 12 or 6) yuan to the platform, he/she could gradually get a consumption rebate of nearly 100 yuan, which was more than 4 (or 8 or 16) times the principal. In less than a month, 5,267 franchisees and 48,505 consumers were involved in the platform. From the project officially opened on December 1, 2016 to the crash at the end of the month, the amount absorbed in just one month reached 1 billion RMB.

Besides ``RenRenGongYi", there are many more platforms of this form and Chinese government have warned the risk of ``consumption rebate" platforms\footnote{http://www.mps.gov.cn/n2253534/n2253543/c6108362/content.html}. There are many other forms of pyramid schemes, such as IGOFX \footnote{\text{https://news.china.com/news100/11038989/20170707/30931598\_all.html}} originated from Malaysia and MMM\footnote{https://en.wikipedia.org/wiki/MMM\_(Ponzi\_scheme\_company)} originated from Russia.

Pyramid schemes are different from ordinary Ponzi schemes named after the eponymous fraudster Charles Ponzi(1882-1949), though in both Ponzi and pyramid schemes, existing investors are paid by the money of new investors\footnote{http://www.51voa.com/VOA\_Special\_English/pyramid-vs-ponzi-69196.html}. In a Ponzi scheme participants believe they are actually earning returns from their investment. While in a pyramid scheme, participants are aware that they are earning money by finding new participants. They become part of the scheme.

Pyramid schemes and Ponzi schemes have been researched from some different perspectives. Joseph Gastwirth proposed a  probability model of a pyramid scheme and concluded that the vast majority of participants have less than a ten percent chance of recouping their initial investment~\cite{Gastwirth1977A}. Stimulated by the Madoff investment scandal in 2008, Marc Artzrouni put forward a first order linear differential equation to describe the
Ponzi schemes~\cite{Artzrouni2009The}. The model of Marc Artzrouni depends on the following parameters: a promised but unrealistic interest rate, the actual realized nominal interest rate, the rate at which new deposits are accumulated and the withdrawal rate. Marc Artzrouni gave the conditions on these parameters for the Ponzi scheme to be solvent or to collapse. The model was fitted to data available on Charles Ponzi's 1920 eponymous scheme and
illustrated with a philanthropic version of the scheme. Tyler Moore et al. make an empirical analysis of these High Yield Investment Programs but not put forward a mathematical model\cite{Moore2012}.  A High Yield Investment Program (HYIP) is considered to be an online Ponzi
scheme, because it pays outrageous levels of interest using money from new investors. Different from the traditional Ponzi schemes, there are
many sophisticated investors understanding the fraud, but hope to profit by joining early, and investors can not withdraw their money at any time.
Anding Zhu, Peihua Fu et al. researched some problems when Ponzi scheme diffuses in complex networks\cite{Zhu2017Ponzi,Fu2017Threshold}.

Since the introduction of random networks, small-world networks and scale-free networks, complex networks have attracted great attention from researchers in various fields such as management and statistical physics. Researches show that many natural and social phenomena have small-world or scale-free characteristics. At present, complex networks have been successfully applied to improve transportation networks\cite{H2015Complex,Liu2015Load}, analyze innovative networks\cite{inproceedings}, research the spread of infectious diseases and rumors\cite{Pastor2000Epidemic,Moreno2004Dynamics}.

Most of the existing literature focuses on the research of Ponzi schemes, including the spread of Ponzi schemes in complex networks, while the research on pyramid scheme is relatively few. To the best of our knowledge, for the pyramid schemes of ``consumption rebate" type, no scholar has put forward a model based on it at present. In order to understand and explain the operation mechanism and characteristics of the pyramid schemes of ``consumption rebate¡± type, then provide ideas for monitoring this kind of pyramid schemes, and offer the basis for further research, we propose a pyramid scheme model which has the principal characters of many pyramid schemes appeared in recent years: promising high returns, rewarding the participants recruiting the next generation of participants, and the organizer will take all the money away when he finds the money from the new participants is not enough to pay the previous participants interest and rewards. We assume the pyramid scheme carries on in the tree network, ER random network, SW small-world network or BA scale-free network respectively, then give the analytical results of how many generations the pyramid scheme can last in these cases. We also use our model to analyse a pyramid scheme in the real world and we find the connections between participants in the pyramid scheme may constitute a SW small-world network.

This paper is organized as follows. In Sec.~2, we briefly introduce the tree network, the random network, the small-world network and the scale-free network. In Sec.~3, we propose our pyramid scheme model. In Sec.~4, we analyse a pyramid scheme in real world. Some discussions and conclusions are given in Sec.~5.

\section{Networks}
\subsection{Tree network}
Tree networks are connected acyclic graph. The word ``tree" suggests branching out from a root and never completing a cycle.
Tree networks are hierarchical, and each node can have an arbitrary number of child nodes. Trees as graphs have many applications, especially
in data storage, searching, and communication~\cite{West2001Introduction}.
\subsection{Random network}
Random network, also known as stochastic network or stochastic graph, refer to complex network created by stochastic process. The most typical random network is the ER model proposed by Paul Erd\"os and Alfred R\'eney~\cite{articleER}. ER model is based on a ``natural" construction method: suppose there are $n$ nodes, and assume that the possibility of connection between each pair of nodes is constant $0 < p < 1$. The network constructed in this way is ER model network. Scientists first used this model to explain real-life networks.
\subsection{Small-world network}
The original model of small-world was first proposed by Watts and Strogatz, and it is the most classical model of small-world network which called SW small-world network~\cite{Watts1998Collective}. The WS small-world network model can be constructed as follows:
take a one-dimensional lattice of L vertices with connections or bonds between nearest neighbors and periodic boundary conditions (the lattice is a ring), then go through each of the bonds in turn and independently with some probability $\phi$ ``rewiring" it. Rewiring in this context means shifting one end of the bond to a new vertex chosen uniformly at random from the whole lattice, with the exception that no two vertices can have more than one bond running between them, and no vertex can be connected by a bond to itself.
The most striking feature of small-world networks is that most nodes are not neighbors of one another, but the neighbors of any given node are likely to be neighbors of each other and most nodes can be reached from every other node by a small number of hops or steps.
It has been found that many networks in real life small-world property, such as social networks~\cite{Grossman2002Reviews}, the connections of neural networks~\cite{Watts1998Collective}, and the bond structure of long macromolecules in the chemical~\cite{Santos2004Topology}.
\subsection{Scale-free network}
 A scale-free network is a network whose degree distribution follows a power law, at least asymptotically. The first model of scale-free network is
proposed by Barabasi and Albert, which is called BA scale-free network~\cite{Barabasi1999Albert}. BA model describes a growing open system starting from a group of core nodes, new nodes are constantly added to the system. The two basic assumptions of BA scale-free network model are: (1) from $m_0$ nodes, a new node is added to each time step, and m nodes are selected to be connected to the new node in $m_0$ nodes($m\leq m_0$); (2) The probability $\Pi_i$ that the new node is connected to an existing node $i$ satisfies $\Pi_i=k_i/\sum^{N-1}_{j=1}k_j$, where $k_i$ denotes the degree of the node $i$ and $N$ denotes the number of nodes. In this way, when added enough new nodes, the network generated by the model will reach a stable evolution state, and then the degree distribution follows the power law distribution. Ref~\cite{Heidelberg2001Statistical} reported that the degree
distribution of many networks in real world is approximate or exact obedience to power law distribution.
\section{The model}
\subsection{Assumptions}
We consider a simple pyramid scheme while it meets the basic features of many pyramid schemes in the real world, especially the ``consumption rebate" platforms. First, it has an organizer that attracts participants through promising high rate of return compared to normal interest rate. Besides the promising return, any participant will be rewarded by the organizer with a proportion of the total investment of the participants he or she directly attracted, thus the early participants will be motivated enough to recruit the next-generation participants and the next-generation participants will do the same thing in order to get more returns. Secondly, we assume all the participants at current generation are recruited by the participants at the upper generation, and the organizer pays the participants at the previous generations the interests and rewards when all possible participants at current generation have joined in the scheme. The third assumption is that the organizer will take all the money away when he finds the money from the new participants is not enough to pay the previous participants interest and rewards. To simplify the model, we also assume all the participants invest the same amount of money and invest only once.
\begin{figure}[htb]
\centering\includegraphics[width=4.2in]{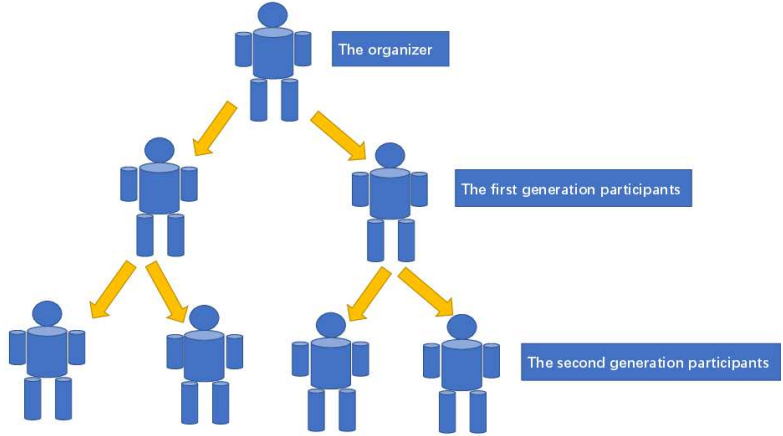}
\caption{A schematic diagram of pyramid scheme. From top to bottom are the organizer, the first generation participants and the second generation participants.}\label{PS}
\end{figure}
Figure~\ref{PS} is a schematic diagram of pyramid scheme, it has one organizer and two generations of participants.

Based on these assumptions, we discuss the pyramid scheme spreads in tree network, random network, small world network, and scale-free network below.
\subsection{Tree network case}
If the pyramid scheme expands in the form of tree network that has a constant branching coefficient $\alpha$ and the root node of the tree network represents the organiser, we can simply write the number of participants at the $g$-th
generation as $n_1\alpha^{g-1}$ and the total amount of money entering the pyramid scheme at the $g$-th generation as $mn_1\alpha^{g-1}$, where $n_1$ is the number of participants at the first generation and $m$ is the money amount of every participant invests. For simplification, we assume $n_1=\alpha$ and $m=1$. We suppose the number of all potential participants is $N$ in this case. Removing the interest and rewards, the relationship between the net inflow of money $M$ of the pyramid scheme and the generation $g$ when all possible participants at $g$-th generation have joined in the scheme can be given by
\begin{equation} \label{1}
M(g)=\alpha^{g} - r_0\sum\limits^{g-1}_{i=1}\alpha^{i}- r_1  \alpha^{g},
\end{equation}
where $r_0$ is the promised rate of return of the organizer, $r_1$ is the ratio of the money rewarded with a participant to the total investment of the participants he or she directly recruited. Normally in real pyramid scheme cases, $r_0$ and $r_1$ are between 0\% and 50\%. The first term of Eq.~({\ref{1}}) represents the investment of all the participants, the second term represents the interest paid to the participants before the generation $g$, and the third term represents the rewards paid to the recruiters of participants at the $g$-th generation.
Notice in our pyramid scheme, the participants at the $g$-th generation are all recruited by the participants at $g-1$-th generation.

The second term of Eq.~({\ref{1}}) is the sum of geometric sequences, after summing them up Eq.~({\ref{1}}) can be rewritten as
\begin{equation}\label{2}
M(g)=\frac{\alpha}{\alpha-1}[(1-r_1)\alpha^g-(1+r_0-r_1)\alpha^{g-1}+r_0].
\end{equation}
Through Eq.~({\ref{2}}) we can find that if the branching coefficient $\alpha$ satisfy the condition
\begin{equation}
\alpha\geq\frac{1-r_1+r_0}{1-r_1},
\end{equation}
the inflow of money $M(g)$ of the pyramid scheme is always positive, so the pyramid scheme will continue forever under the circumstances.

However, the potential participants are limited to $N$ and the pyramid scheme will stop eventually. The maximum generation $G$ of the pyramid scheme is given by
\begin{equation}
G_{TR} = \lfloor\log_\alpha(\frac{N\alpha-N+\alpha}{\alpha})\rfloor + 1,
\end{equation}
where $\lfloor x\rfloor$ is the integer part of $x$. At the $G$-th generation all the potential participants have joined the pyramid scheme, and the
organizer will take away all the money and don't pay the interest and rewards any more. We can write the final income of the pyramid as
\begin{equation}
R_p=N-r_0\sum\limits^{G-2}_{i=1}(G-i-1)\alpha^{i}-r_1\sum\limits^{G-1}_{i=2} \alpha^{i},
\end{equation}
and the income of the participants at $i$-th generation is
\begin{equation} \label{income}
R_i=\left\{
\begin{array}{rcl}
&r_0(G-i-1)\alpha^{i}+r_1 \alpha^{i+1}-\alpha^{i}, & {\rm for}\ \  1\leq i \leq G-2. \\
&-\alpha^{i}, & {\rm for}\ \  G-2<i \leq G.
\end{array} \right.
\end{equation}
\begin{figure}[htbp]
\begin{center}
\subfigure[]
 {\includegraphics[ width = 3.0in]{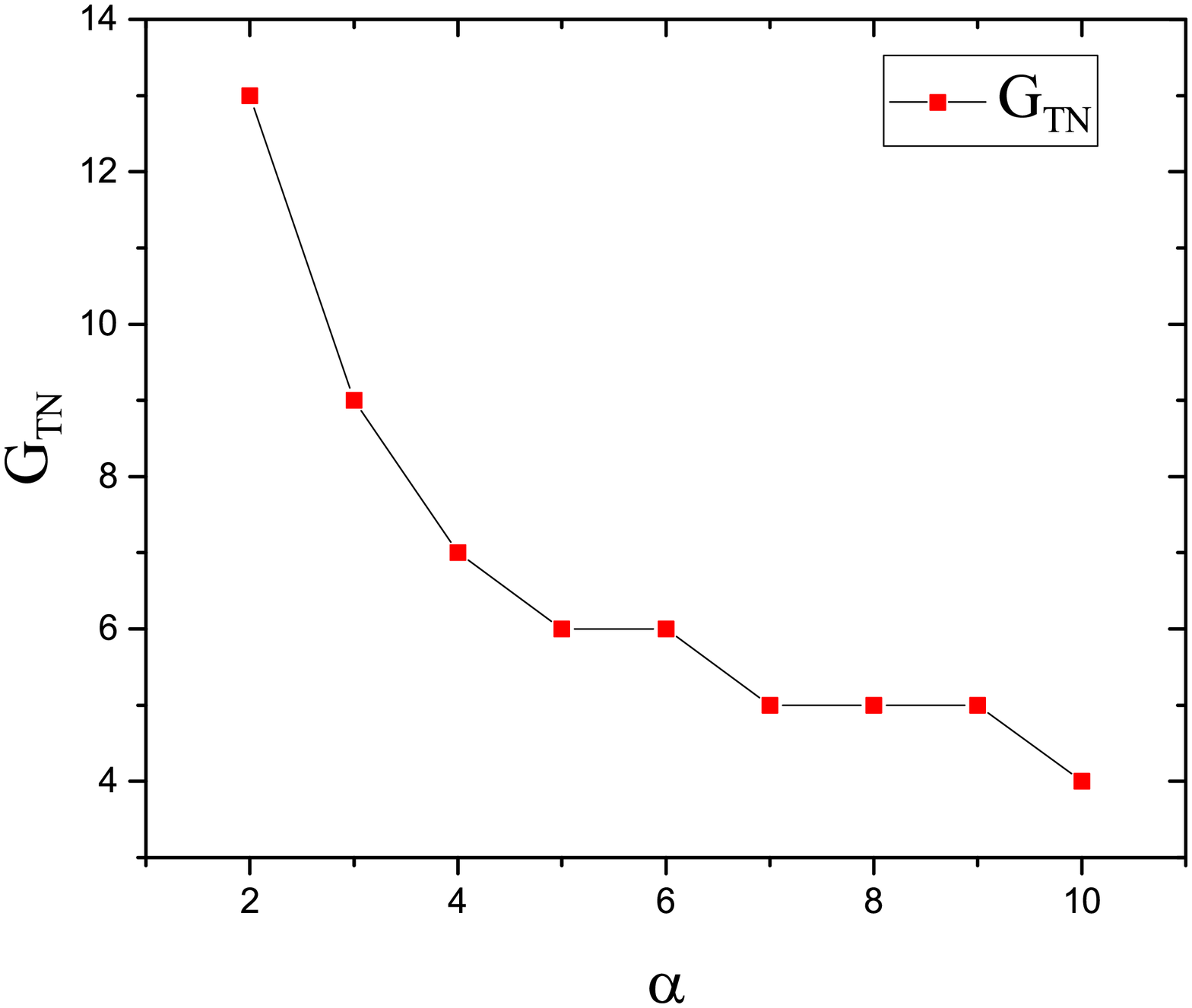}
 \label{fig1}}
\hspace{0\textwidth}
\subfigure[]
 {\includegraphics[ width = 3.0in]{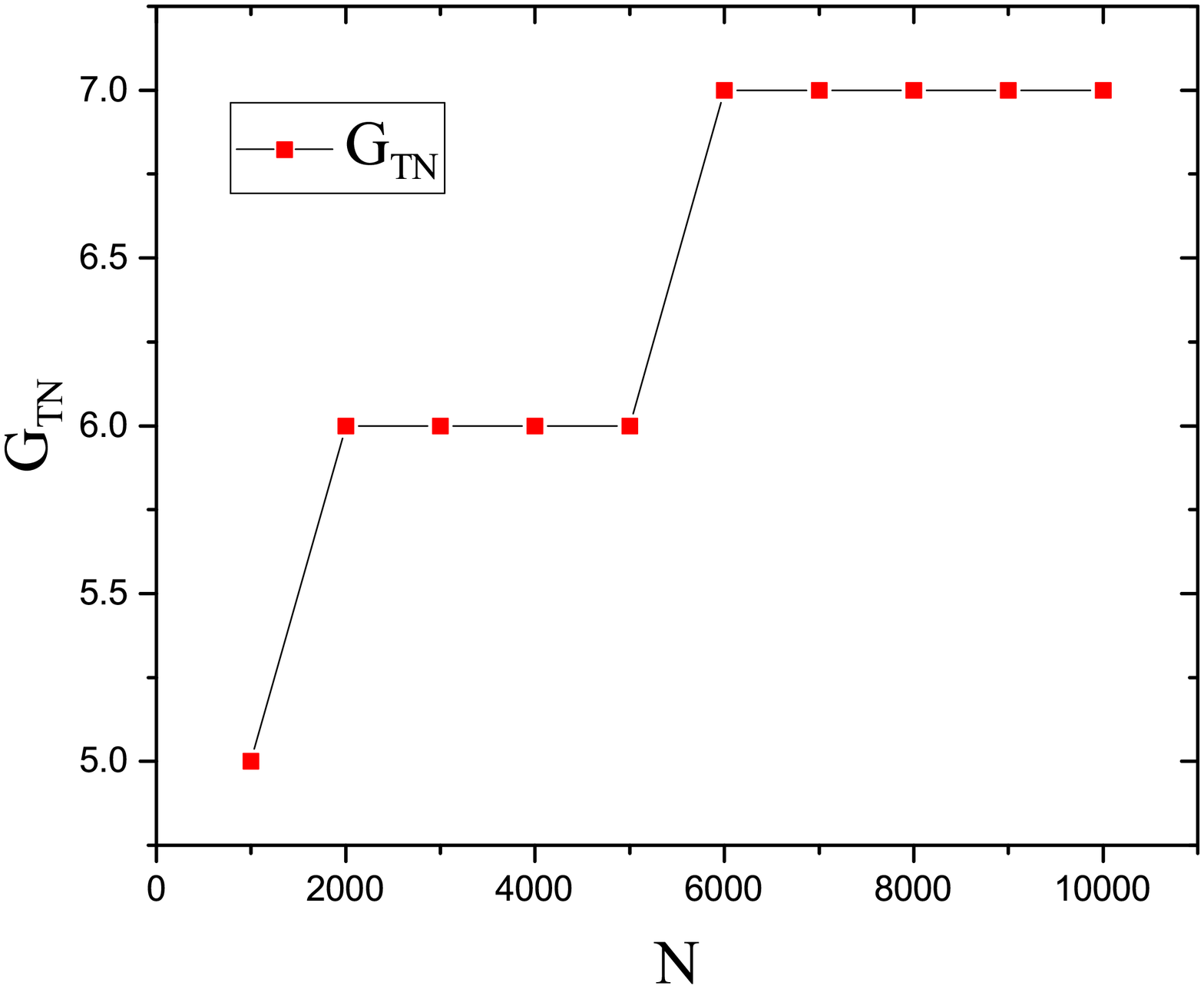}
 \label{fig2}}
\caption{(a) The analytical result and the simulative result of maximum generation $G_{ER}$ when the branching coefficient $\alpha$ changes. We take the parameters value $N=10000$, $r_0=0.1$, $r_1=0.1$; (b) The analytical result and the simulative result of maximum generation $G_{ER}$ when the possible participants $N$ changes. We take the parameters value $\alpha=4$, $r_0=0.1$, $r_1=0.1$.
} \label{tn}
\end{center}
\end{figure}

Figure~\ref{tn}(a) shows the analytical result and the simulative result of maximum generation $G_{ER}$ when the branching coefficient $\alpha$ changes, and we take the parameters value $N=10000$, $r_0=0.1$, $r_1=0.1$. Figure~\ref{tn}(b) shows the analytical result and the simulative result of maximum generation $G_{ER}$ when the possible participants $N$ changes, and we take the parameters value $\alpha=4$, $r_0=0.1$, $r_1=0.1$. Figure~\ref{tn} illustrates intuitively that in the tree network case, if other conditions of the pyramid scheme remain unchanged, the larger the branch coefficient, that is, the more new participants each person recruits, the fewer generations the pyramid scheme can last. On the other hand, when other conditions remain unchanged, the larger the number of potential participants, the more generations the pyramid scheme can sustain, but every new generation needs more participants, and this growth of new participants is exponential.

\subsection{Random network case}
If the pyramid scheme takes place in a ER random network has an average degree $k$ and $N$ nodes, we assume the organizer is a random node in the network and other nodes represent the potential participants. The organizer recruits the potential participants nearest to him as the first generation participants, and the first generation participants recruits the potential participants nearest to them as the second generation participants, and so on. So the value of the generation of any participant in the pyramid scheme is the shortest path length from the node represents the organizer. Ref.~\cite{article:Katzav} gives the approximate analytical results for the distribution of shortest path lengths in ER random networks, the number of nodes at the $i$-th generation is about $k^{i}$ if $i*log_Nk<1$ and all the nodes are included in the pyramid scheme if $i*log_Nk>1$. Therefore, the pyramid scheme in ER random network is approximate to the case in the tree network above and the difference is the branching coefficient $\alpha$ should be replaced by the average degree $k$.

\begin{figure}[htbp]
\begin{center}
\subfigure[]
 {\includegraphics[ width = 3.0in]{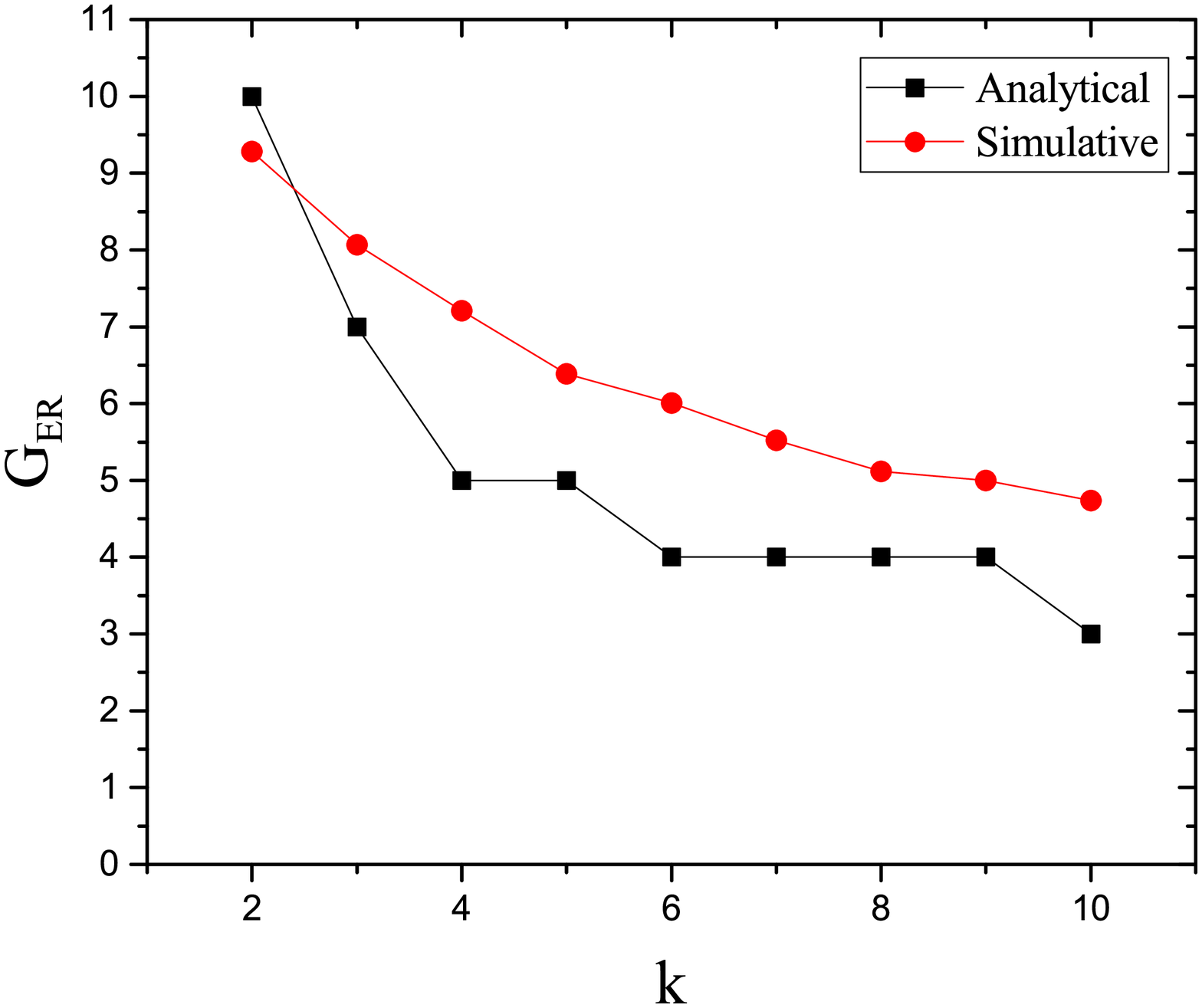}
 \label{fig1}}
\hspace{0\textwidth}
\subfigure[]
 {\includegraphics[ width = 3.0in]{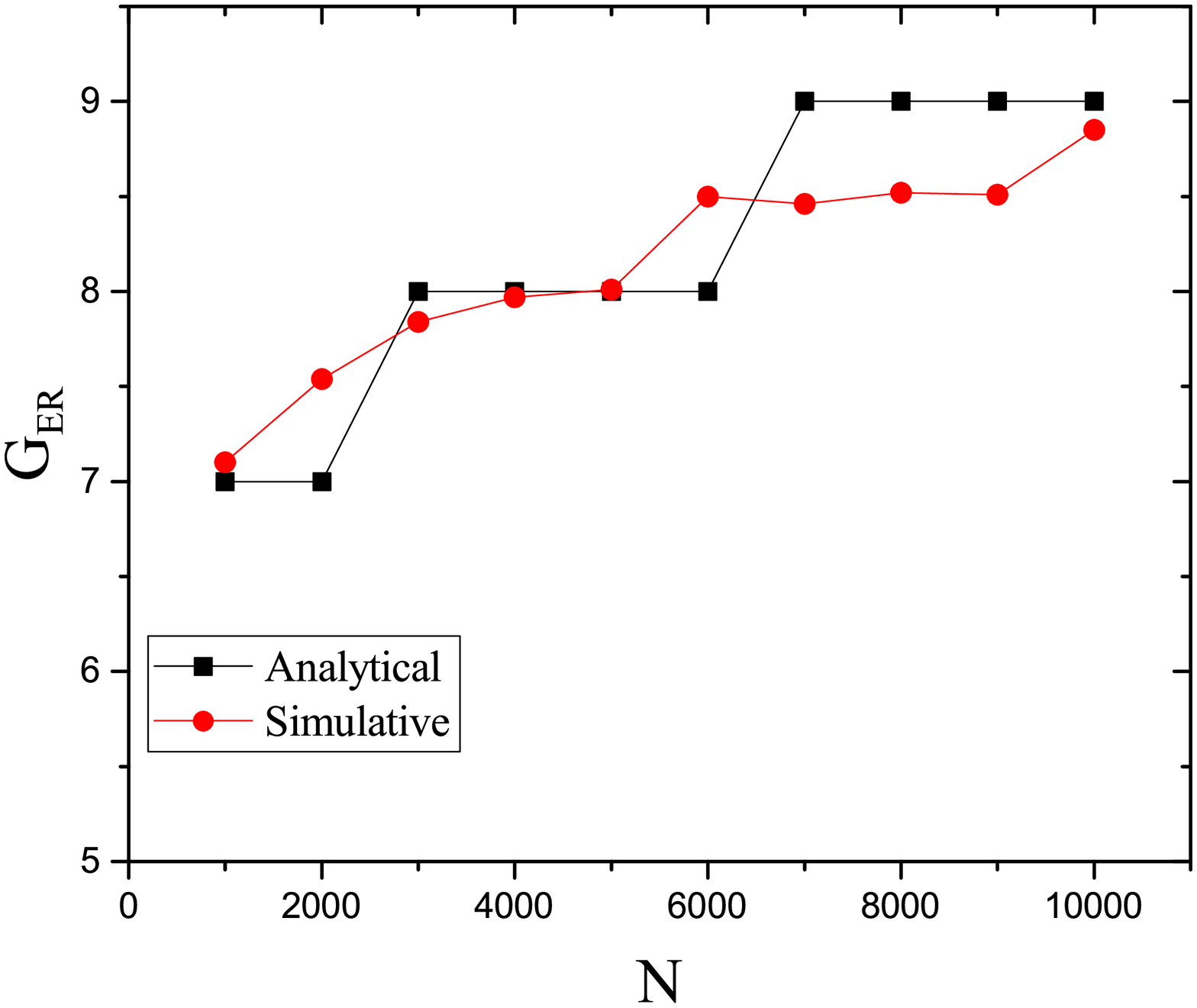}
 \label{fig2}}
\caption{(a) The analytical result and the simulative result of maximum generation $G_{ER}$ when the average degree $k$ changes. We take the parameters value $N=1000$, $r_0=0.1$, $r_1=0.1$; (b) The analytical result and the simulative result of maximum generation $G_{ER}$ when the possible participants $N$ changes. We take the parameters value $k=4$, $r_0=0.1$, $r_1=0.1$. The simulative results in the (a) and (b) are averaged after 100 simulations.
} \label{er}
\end{center}
\end{figure}

Firstly, like the case in tree network, $r_0$, $r_1$ and $k$ should satisfy the following condition:
\begin{equation}\label{cer}
k\geq\frac{1-r_1+r_0}{1-r_1}.
\end{equation}
The approximate maximum generation $G$ of the pyramid scheme in this case is given by
\begin{equation} \label{rm}
G_{ER} \approx \lfloor 1/  \log_Nk\rfloor+1.
\end{equation}
In addition, we can also write the approximate expressions of the organiser's and participants' income which have the same form of Eq.~({\ref{income}}), which we omits it here.
Figure~\ref{er}(a) The analytical result and the simulative result of maximum generation $G_{ER}$ when the average degree $k$ changes, and we take the parameters value $N=1000$, $r_0=0.1$, $r_1=0.1$. Figure~\ref{er}(b) shows the analytical result and the simulative result of maximum generation $G_{ER}$ when the possible participants $N$ changes, and we take the parameters value $k=4$, $r_0=0.1$, $r_1=0.1$. The simulative results in the (a) and (b) are averaged after 100 simulations. From Figure~\ref{er}, we can find that in ER random network case, the relationship between maximum generation $G_{ER}$ and mean degree $k$, and the relationship between $G_{ER}$ and potential participants $N$ are similar to the case in tree network, where the mean degree $k$ represents the amount of participants that each participant can recruit averagely. We can also find that within the range of parameters we have chosen, analytical results and simulative results are very close.
\subsection{Small world network case}
Now we consider the pyramid scheme carries on in a SW small-world network, to some extent this case is similar to the case in ER random network. We also randomly choose a node as the organiser, other nodes represent the potential participants, and $r_0$, $r_1$ represent interest rate and reward ratio respectively. The value of the generation of any participant in the pyramid scheme is the shortest path length from the node represents the organizer. Ref.~\cite{Newman1999Scaling} points out that the number of nodes increases exponentially with the average length of the shortest path when the nodes are infinite.
The approximate surface area of a sphere of radius $r$ on the SW small-world network can be given by\cite{Newman1999Scaling}
\begin{equation}
A(r) = 2 e^{4r/\xi},
\end{equation}
where $\xi=1/\phi k$, and $\phi$ is rewiring probability and $k$ is the degree of the corresponding rule graph.

Change $r$ to $g$, we can obtain the approximate number of participants at $g$-th generation. Because of the exponential form of $A(g)$,
we can deal with this case just like in the cases of tree network and ER random network. The branching coefficient $\alpha$ should be replaced by $e^{4/\xi}$, and the following condition should be satisfied:
\begin{equation} \label{swc}
e^{4/\xi}\geq\frac{1-r_1+r_0}{1-r_1}.
\end{equation}

If the nodes are finite, the number of nodes reach the peak when the distant from the node to the organiser is near the average length of the shortest path. If greater than the average length of the shortest path, numbers are quickly reduce to 0, so it can be approximately considered that the maximum generation $G$ is close to the value of the average length of the shortest path.
The average path length $\overline{d}$ of the SW small-world network is given by~\cite{Newman1999Scaling}
\begin{equation}
\overline{l}_{SW}\approx \frac{N}{K}f(\phi KN),
\end{equation}
where
\begin{equation}
f(u)=\left\{
\begin{array}{rcl}
&\frac{1}{4},& {\rm if}\  u \to 0.  \\
&\ln u/u,& {\rm if}\  u \to \infty.
\end{array}\right.
\end{equation}
The number of nodes with average shortest path length from the node representing the organizer is the largest. So we can infer the maximum generation $G_{SW}$ of the pyramid scheme is given by~\cite{article:A1999On}
\begin{equation}
G_{SW} \approx \lfloor\overline{l}_{SW}\rfloor+1.
\end{equation}

\begin{figure}[h]
\begin{center}
\subfigure[]
 {\includegraphics[ width = 3.0in]{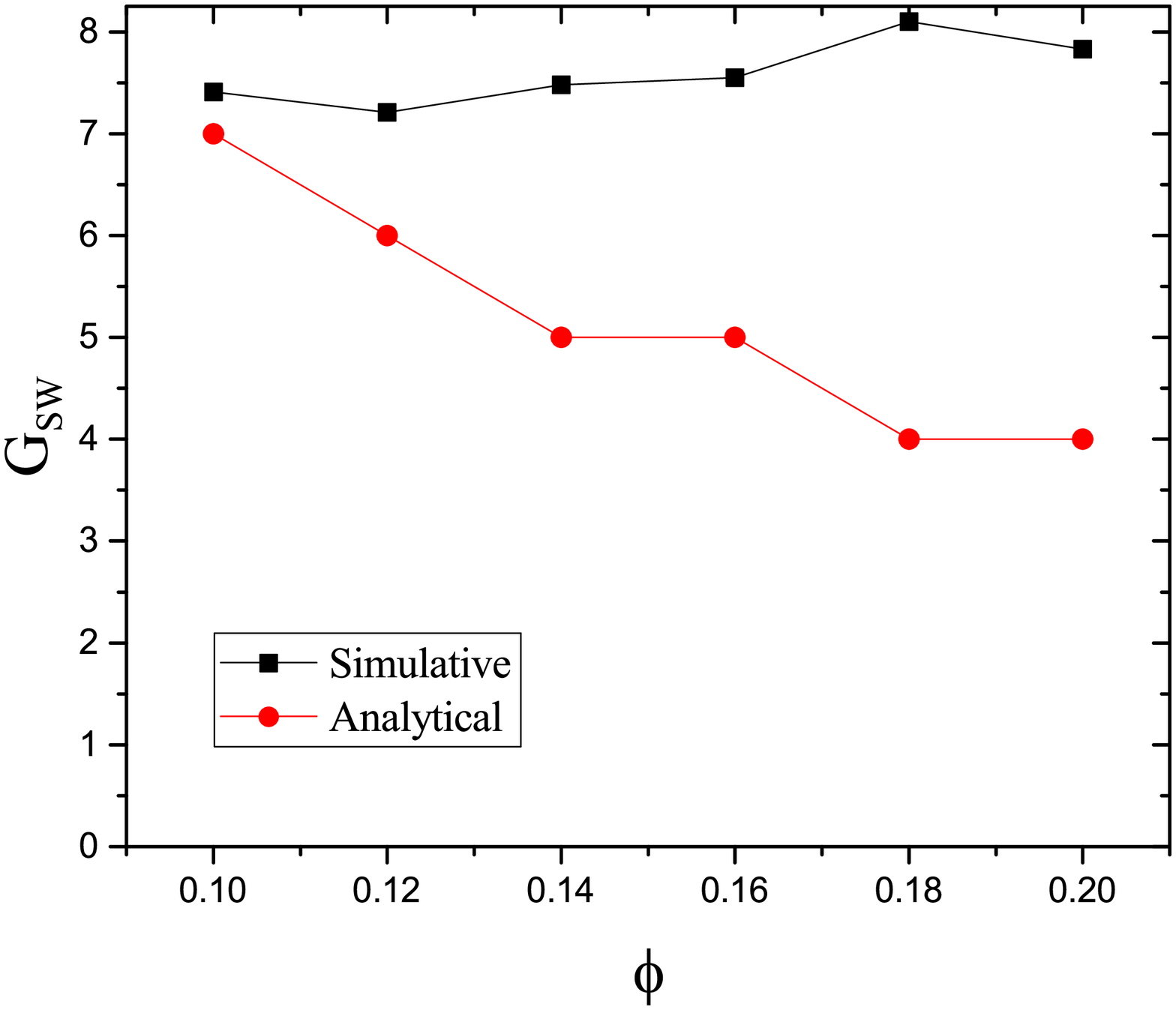}
 \label{fig1}}
\hspace{0\textwidth}
\subfigure[]
 {\includegraphics[ width = 3.0in]{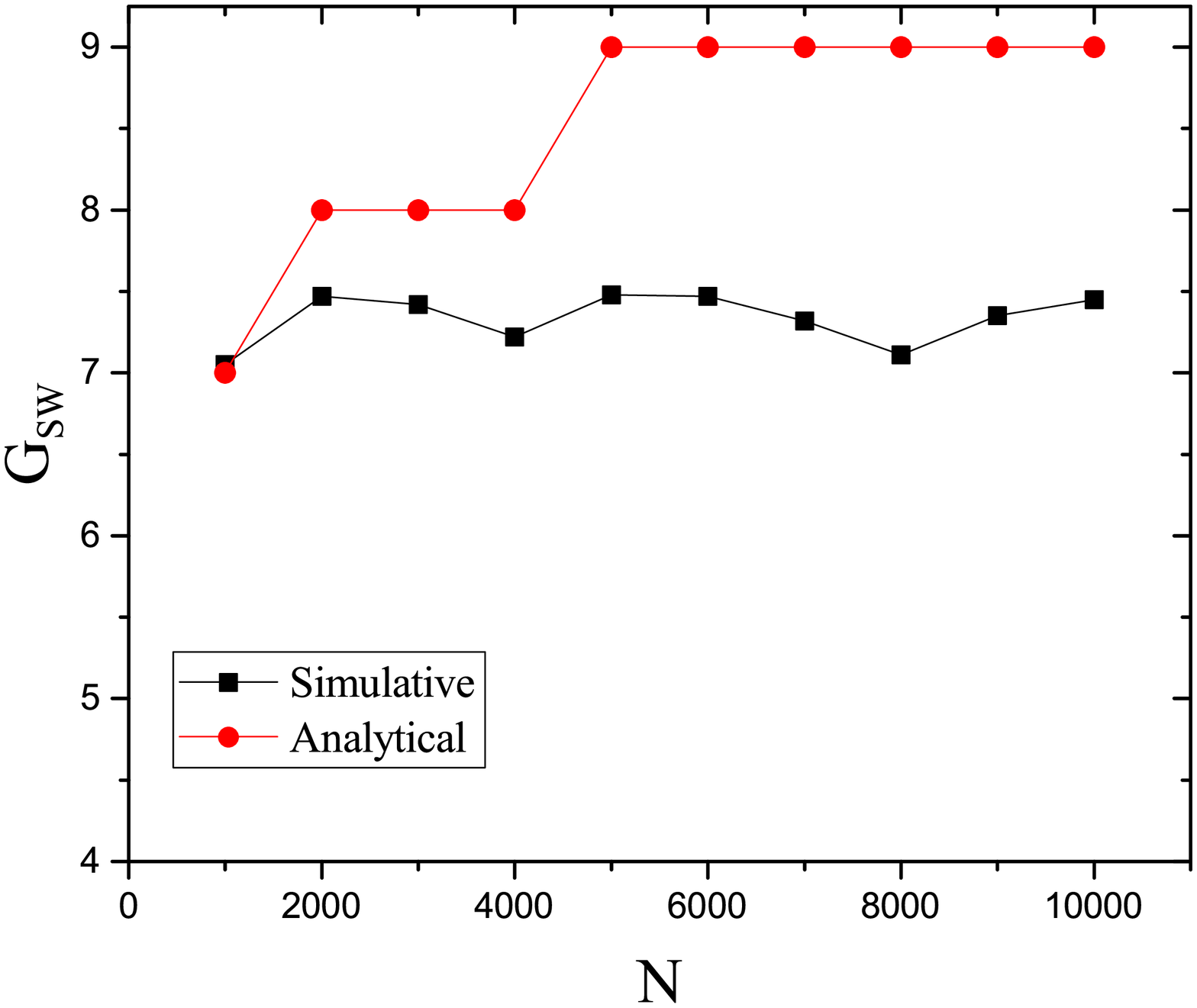}
 \label{fig2}}
\caption{(a) The analytical result and the simulative result of maximum generation $G_{SW}$ when the possible participants $\phi$ changes. We take the parameters value $N=1000$, $K=3$, $r_0=0.2$, $r_1=0.2$; (b) The analytical result and the simulative result of maximum generation $G_{SW}$ when the possible participants $N$ changes. We take the parameters value $K=3$, $\phi=0.1$, $r_0=0.2$, $r_1=0.2$. The simulative results in the (a) and (b) are averaged after 100 simulations.
} \label{sw}
\end{center}
\end{figure}

In the simulation, we find that the values of $r_0$ and $r_1$ are very important. Generally speaking, the greater the values of $r_0$ and $r_1$ satisfy the Eq.~(\ref{swc}), the closer the simulation results and numerical results are. This is because when the values of $r_0$ and $r_1$ are larger, the pyramid scheme can easily terminate when the number of generations exceeds the value of the average shortest path length.
Figure~\ref{sw}(a) shows the analytical result and the simulative result of maximum generation $G_{SW}$ when the possible participants $\phi$ changes, and we take the parameters value $N=1000$, $K=3$, $r_0=0.2$, $r_1=0.2$. Figure~\ref{sw}(b) shows the analytical result and the simulative result of maximum generation $G_{SW}$ when the possible participants $N$ changes, and we take the parameters value $K=3$, $\phi=0.1$, $r_0=0.2$, $r_1=0.2$. The simulative results in the (a) and (b) are averaged after 100 simulations. In Figure~\ref{sw}, we find that within the range of parameters we selected, the maximum generation $G_{SW}$ of the pyramid scheme is not very sensitive to the reconnection probability $\phi$ and the potential participants, and the analytical results are basically in accordance with the simulative results.
\subsection{Scale-free network case}
If the pyramid scheme expands in a BA scale-free network, similar to the cases in ER random network and SW small-world network above, we also randomly choose a node as the organiser and and other nodes represent the potential participants. The organizer recruits participants and the participants recruit the next generation participants through the network connections. To ensure positive inflows, the following condition must be satisfied:
\begin{equation}
(1-r_1)n(g+1)\geq r_0\sum\limits^{g}_{i=1}n(g),
\end{equation}
Where $n(g)$ represents the number of participants at $g$-th generation, and $n(g+1)$ represents the number of participants at $g+1$-th generation.
The distribution of shortest path length approximates the normal distribution and the position corresponding to the highest point of normal distribution is the average shortest path length~\cite{article:Ventrella2018On}. The average path length $\overline{s}$ of the BA scale-free network is given by\cite{article:Reuven2003Scale}
\begin{equation}\label{bac}
\overline{l}_{BA} \approx  \frac{\ln N}{\ln \ln N}.
\end{equation}

\begin{figure}[htbp]
\centering\includegraphics[width=4.2in]{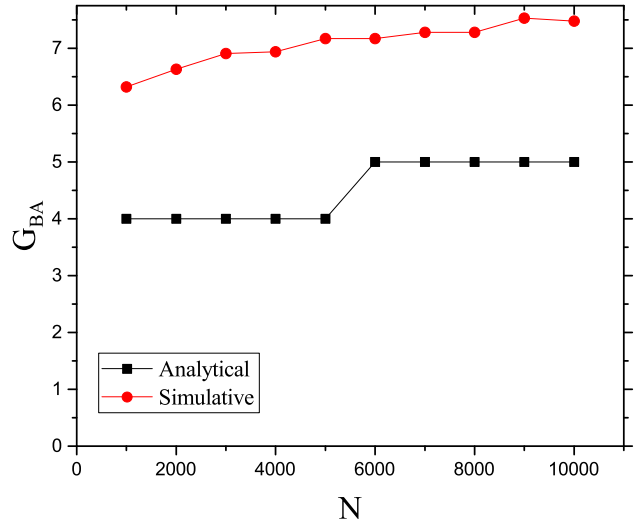}
\caption{The analytical result and the simulative result of maximum generation $G_{BA}$ when the possible participants $N$ changes. We take the parameters value
$r_0=0.2$, $r_1=0.2$. The simulative result is averaged after 100 simulations.}  \label{ba}
\end{figure}

Before the peak, the number of participants per generation grew faster than exponential growth. But after that, The number of participants per generation declined rapidly, so the condition can not be satisfied any more. So we can infer the maximum generation $G_{BA}$ is close to the value of the average shortest path length,
and is given by
\begin{equation}
G_{BA} \approx  \lfloor\overline{l}_{BA}\rfloor +1.
\end{equation}

Figure~\ref{ba} shows the analytical result and the simulative result of maximum generation $G_{BA}$ when the possible participants $N$ changes. We taken the parameters value $r_0=0.2$, $r_1=0.2$, and the simulative result is averaged after 100 simulations. The analytical results can basically reflect this characteristic. We can find that in scale-free networks, the maximum generation $G_{BA}$ is not very sensitive to the potential participants in Figure~\ref{ba}. The analytical results can basically reflect this characteristic.
\section{A pyramid scheme in real world}
Although real cases of pyramid scheme are easy to find in news reports, there are few cases that give details of the number of people involved and the pyramid generations. Usually, when the organizer of the pyramid scheme disappears, the participants with loss will report the case to the police, and the police will investigate the case. On July 23, 2018, China News Network Guangzhou Station reported a pyramid scheme that had 75663 account numbers and 46 generations, and the pyramid scheme had amassed 76 million yuan in three months\footnote{http://www.gd.chinanews.com/2018/2018-07-24/2/397998.shtml}. This is the same type of pyramid scheme as described in the introduction. Using the analysis in Section 3, we assume the pyramid scheme carries on tree network, ER random network, SW small-world network and BA scale-free network respectively, then verify which network can describe the pyramid scheme in the real world well. We assume one account number represent a participant.

If this real pyramid scheme expands in a tree network, we can calculate the tree network' branching coefficient $\alpha \approx 1.28$. This means on average, less then two participants are recruited by each participant. But we can not know more about the connections between the participants except branching coefficient.

If this real pyramid scheme spreads in ER Random network, we can calculate the average degree $k \approx 1.28$ through Eq.~(\ref{rm}). So each node is connected to 1.28 nodes on average, and the connection probability in ER random network is less then $1.28/75663 \approx 1.7\times 10^{-5}$, which is very small then isolated nodes and nodes with degree 1 are easy to appear in the network. Although this case is similar to that of tree network, the branching coefficient in random network is not stable and it's easy to end the pyramid scheme if Eq.~(\ref{ctr}) is not satisfied (the minimum value of the formula $(1-r_1+r_0)/(1-r_1)$ is greater than 1). So we think the pyramid scheme can hardly happen in the ER random network.

If this real pyramid scheme carries on in a BA scale-free network, through the analysis and simulation, we find developing to 46 generations need far more than 75663 participants. Therefore, the connections between participants are impossible to form a BA scale-free network.

\begin{figure}[h]
\begin{center}
\subfigure[]
 {\includegraphics[ width = 3.0in]{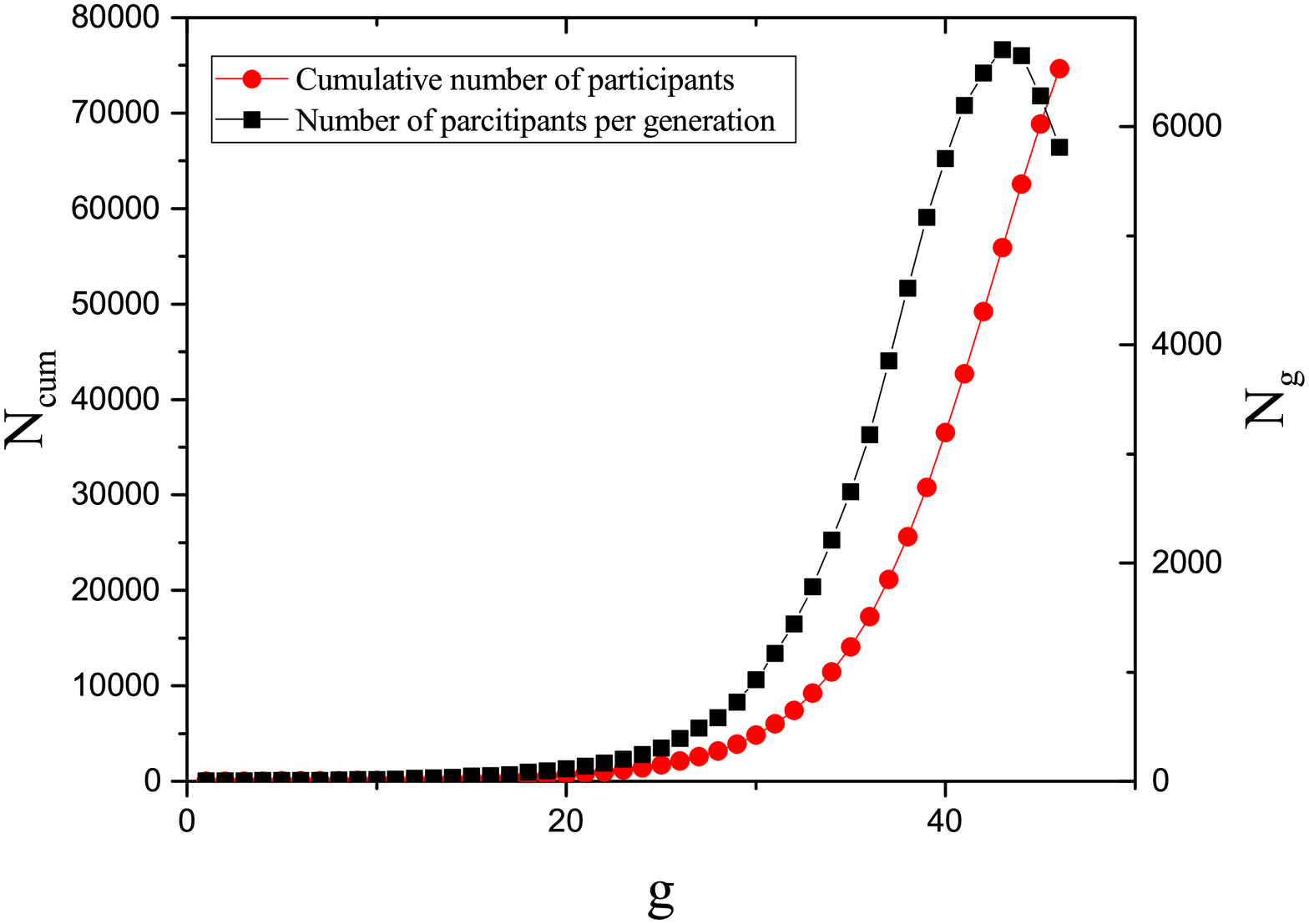}
 \label{fig1}}
\hspace{0\textwidth}
\subfigure[]
 {\includegraphics[ width = 3.0in]{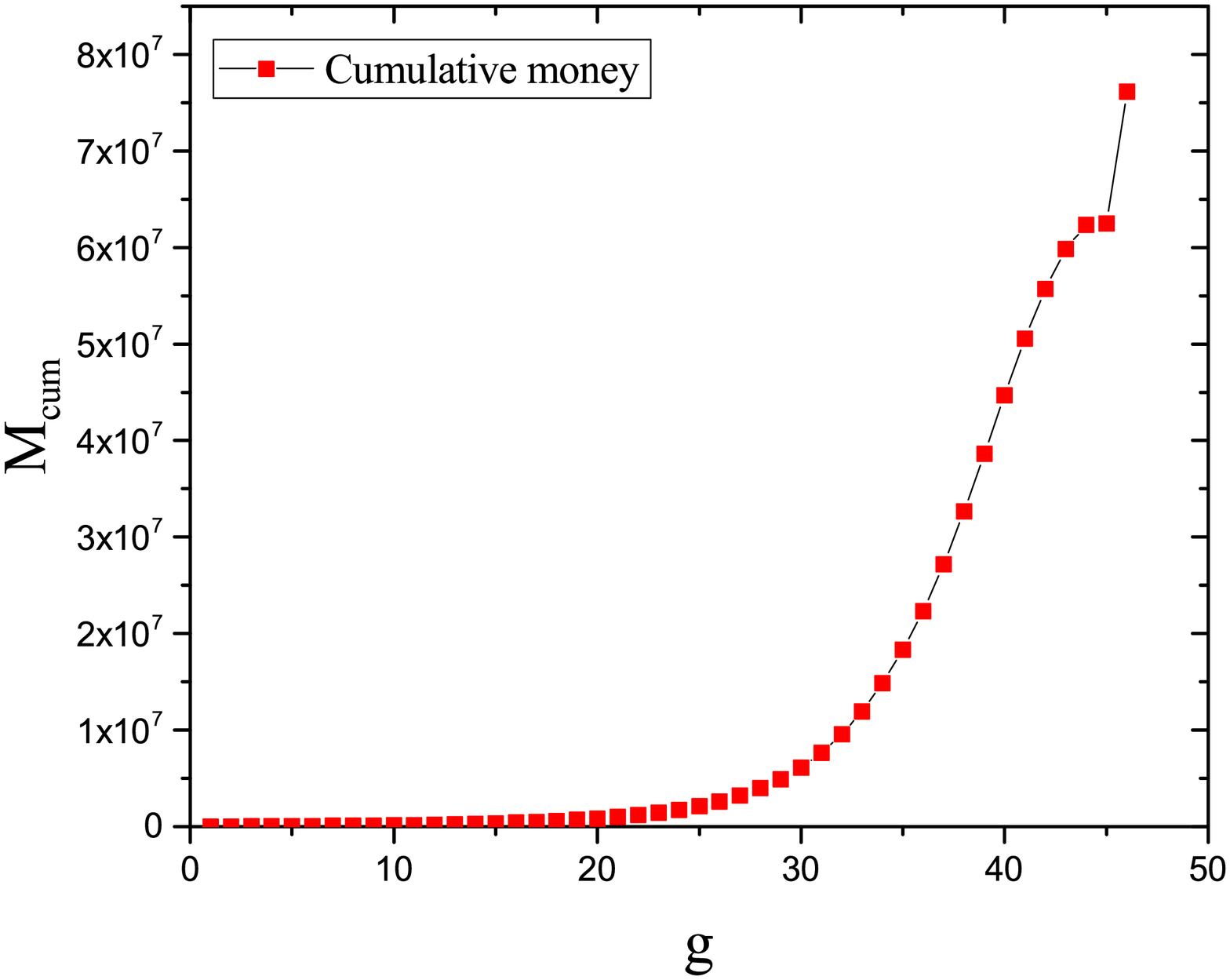}
 \label{fig2}}
\caption{The cumulative number $N_{cum}$ of participants, the number of participants $N_g$ in each generation, and the cumulative money $M_{cum}$ changes over generation $g$ in the simulatve pyramid scheme. This is one simulative result in the SW small-world case which the parameters we select are $N=100000$, $\phi=0.02$, $r_0$=0.1, $r_1=0.1$, and each participant invest 23500 yuan.
}\label{rwe}
\end{center}
\end{figure}

If this real pyramid scheme takes place in a SW small-world network and accords all our assumptions, we could find a simulative result to fit the result of
the real pyramid scheme. The parameters we select are $N=100000$, $\phi=0.02$, $K=4$, $r_0$=0.1, $r_1=0.1$, and each participant invest 23500 yuan. The simulative pyramid scheme has 74652 participants, and develops to 46 generations, and the fund pool of the pyramid is about 76 million yuan. The simulation results are in good agreement with the real pyramid scheme. Figure~\ref{rwe}(a) and~\ref{rwe}(b) shows the cumulative number $N_{cum}$ of participants, the number of participants $N_g$ in each generation, and the cumulative money $M_{cum}$ changes over generation $g$ in the simulative pyramid scheme.

Figure~\ref{rwe} shows that, the amount of participants and the amount of accumulated money of the pyramid scheme grow slowly in the initial stage and explosively in the later stage. Once the growth rate slows down, the amount of the pyramid scheme's accumulated money will reach a peak soon and the organizer will escape. The probability of reconnection in simulation is 0.02, which can be understood according to the actual situation and means that participants tend to recruit new participants from familiar people. In fact, according to our investigation and many news reports, such pyramid frauds always arise in small cities, and most of the participants recruit new participants from their familiar people. As generations go on, the network constituted by all participants has the properties of small world: agglomeration and having some flocks, which is similar to the interpersonal network. Although our model has been simplified and approximated, it is enlightening to explain the real case.

Through the above simulation analysis, we can speculate that the connections between participants in the real case may constitute a SW small-world network.

\section{Conclusion}
In summary, we have proposed a pyramid scheme model which has the principal characters of many pyramid schemes appeared in recent years: promising high returns and rewarding the participants attracting the next generation of participants. Assuming the pyramid scheme spreads in the tree network, ER random network, SW small-world network and BA scale-free network, we give the conditions for the continuity of the pyramid scheme, and the analytical results of how many generations the pyramid scheme can last if the organizer of the pyramid scheme takes all the money away when he finds the new money is not enough to pay interest and incentives. We also use our model to analyse a pyramid scheme in the real world and the result displays the connections of  participants in the pyramid may have the feature of small world.

Our work is helpful to understand the operation mechanism and characteristics of the pyramid schemes of ``consumption rebate" type. Our model may be able to apply to some current illegal high-interest loans, if these illegal projects promise a high interest rate and reward the investors who encourage others to invest in such projects, but the money accumulated is not actually invested in any real projects. Our work shows that the pyramid schemes of
``consumption rebate" type are not easy to be detected by the supervision because of the small amount of funds and small amount of participants accumulated in the initial stage. After the rapid growth of funds and participants, it often comes to the end of this kind of pyramid frauds, and the organizers often have already fled. So for regulators, it's better to nip such platforms in the bud so that more people don't suffer loss. In addition, to some extent, our work provides some basis for further study of such frauds. For example, we will further consider how the participants' beliefs about always having enough new participants effect the operation of such frauds.
\section{Acknowledgement}
This research was supported by National Natural Science Foundation of China (No.71771204, 91546201) .

\bibliographystyle{ieeetr}
\bibliography{reference}
\end{document}